\documentclass[%
 reprint,
superscriptaddress,
longbibliography,
 amsmath,amssymb,
 aps,
pra
]{revtex4-2}

\usepackage{graphicx}
\usepackage{dcolumn}
\usepackage{xcolor}
\usepackage{bm}
\usepackage{hyperref}
\usepackage{amsmath}

\begin{document}

\preprint{APS/123-QED}

\title{Mott-insulating phases of the Bose-Hubbard model on quasi-1D ladder lattices}

\author{Lorenzo Carfora}
\email{lorenzo.carfora.2018@uni.strath.ac.uk}
\affiliation{Department of Physics and SUPA, University of Strathclyde, Glasgow G4 0NG, United Kingdom}
\author{Callum W. Duncan}
\affiliation{Aegiq Ltd., Cooper Buildings, Arundel Street, Sheffield S1 2NS, United Kingdom.}
\affiliation{Department of Physics and SUPA, University of Strathclyde, Glasgow G4 0NG, United Kingdom}
\author{Stefan Kuhr}%
\affiliation{Department of Physics and SUPA, University of Strathclyde, Glasgow G4 0NG, United Kingdom}
\author{Peter Kirton}%
\email{peter.kirton@strath.ac.uk}
\affiliation{Department of Physics and SUPA, University of Strathclyde, Glasgow G4 0NG, United Kingdom}

\date{\today}

\begin{abstract} 

We calculate the phase diagram of the Bose-Hubbard model on a half-filled ladder lattice including the effect of finite on-site interactions. This shows that the rung-Mott insulator (RMI) phase persists to finite interaction strength, and we calculate the RMI-superfluid phase boundary in the thermodynamic limit. We show that the phases can still be distinguished using the number and parity variances, which are observables accessible in a quantum-gas microscope. Phases analogous to the RMI were found to exist in other quasi-1D lattice structures, with the lattice connectivity modifying the phase boundaries.
This shows that the the presence of these phases is the result of states with one-dimensional structures being mapped onto higher dimensional systems, driven by the reduction of hopping rates along different directions.

\end{abstract}

\maketitle


\newcommand{\callum}[1]{\textcolor{blue}{[#1]}}
\newcommand{\lorenzo}[1]{\textcolor{red}{[#1]}}
\newcommand{\peter}[1]{\textcolor{pink}{[#1]}}

\definecolor{fig_green}{RGB}{50, 205, 50}
\definecolor{fig_red}{RGB}{255, 0, 0}
\definecolor{fig_purple}{RGB}{128, 0, 128}

\section{\label{sec:level1} Introduction}





Lattice dimensionality and geometry have a profound impact on the physical properties of strongly-correlated quantum systems. 
In ultracold-atom experiments, control over the geometry of the lattice~\cite{stoferle2004transition, bloch2008many, gross2017quantum} has proved instrumental in the investigation of emergent many-body phenomena such as frustration~\cite{jo2012ultracold, tarruell2012creating}, non-trivial band structures~\cite{aidelsburger2013realization, atala2013direct, meier2016observation, leder2016real} and dimensional crossovers~\cite{revelle20161d, yao2023strongly, guo2024observation}. These studies show that ultracold atoms in optical lattices are an ideal platform to demonstrate how dimensionality and lattice geometry affect complex quantum phenomena.

Among these engineered geometries, 
the ladder lattice has proved to be a useful platform for investigating low-dimensional quantum systems~\cite{cazalilla2004bosonizing, cazalilla2011one, giamarchi2016one}. Its quasi-1D structure enables the observation of a crossover from 1D to 2D physics~\cite{vogler2014dimensional}, which is especially evident in its response to a magnetic field. For example, the phase transitions between Meissner and vortex phases were both experimentally~\cite{atala2014observation} and theoretically~\cite{commensurate_ladder_giamarchi, tokuno2014ground, piraud2015vortex, Buser2020Interacting} observed in ladder lattices. 
These flux-driven transitions produce rich phase diagrams due to their interplay with population-filling-driven transitions~\cite{petrescu2015chiral, DiDio2015persisting, orignac2016incommensurate, impertro2025strongly}. Flux-driven and filling-driven transitions share analogous underlying mechanisms~\cite{Orignac2001Meissner, giamarchi1997Mott}.  This is because both are related to the commensurability of the states involved. In particular, even in the absence of a magnetic field, the ladder geometry allows for additional commensurate fillings and associated transitions. While the ground state of 1D lattices is a Mott-insulator (MI) only at integer fillings~\cite{greiner2002quantum,jaksch1998cold, kuhner1998phases, bloch2008many, krutitsky2016ultracold, di2024commensurate}, the ladder supports insulating phases at both integer~\cite{luthra2008phase, commensurate_ladder_giamarchi} and half-integer fillings~\cite{crepin2011phase}. The insulating phase which appears  at half-integer fillings, induced by strong coupling between the chains, is known as the rung-Mott insulator (RMI).




The critical behavior of ladder systems has been studied via bosonization and renormalization group techniques~\cite{giamarchi1997Mott, Orignac2001Meissner, Petrescu2017precursor}. Most of the previous studies of the RMI focused on the hardcore boson (HCB) limit as a way to uncover the differences between Bose and Fermi statistics. While the mapping of HCBs onto spinless fermions is exact in a 1D chain~\cite{girardeau1960relationship}, it breaks down in a ladder once inter‑chain couplings are introduced~\cite{crepin2011phase}. This gives rise to significant nonlocal interactions, described by complex Hamiltonian terms~\cite{carrasquilla2011bose}. Beyond this, the extension of the RMI phase into the softcore regime has been investigated by considering three‑body local constraints both for attractive and repulsive interactions~\cite{Padhan2023Quantum, singh2014quantum}. Cluster mean‑field and Gross–Pitaevskii approaches have likewise been applied to ladders in the presence of synthetic gauge fields in both the strong and weak magnetic flux limits~\cite{tokuno2014ground, kelecs2015mott}. 
Although these approximations give a comprehensive picture of the system behavior, they may not reliably capture the nuances of the RMI in more experimentally accessible regimes. 

In this work, we investigate the transition between the superfluid (SF) and RMI phases at half-filling and in the presence of finite on-site interactions. We characterize the system using microscopic observables that are accessible experimentally.
We extend the discussion of the RMI phase to analogous insulating phases on different quasi-1D lattices, underlining the connection between filling fraction and dimensionality by relating it to less trivial geometries, such as staggered two-dimensional~\cite{dong2025observation} and triangular~\cite{becker2010ultracold} lattices. We also provide guidance for future experiments by considering quantities which are directly measurable in quantum-gas microscope experiments with site-resolved in-situ imaging of atoms~\cite{bakr2009quantum, sherson2010single, edge2015imaging, haller2015single, kuhr2016quantum, gross2021quantum}. In these setups, optical digital mirror devices allow for the design and microscopic control of the lattice geometry~\cite{preiss2015quantum, gauthier2016direct, zupancic2016ultra}, dimensionality and filling~\cite{choi2016exploring, di2024commensurate} via engineered light potentials. Two-leg ladder systems have also been realized using optical superlattices~\cite{aidelsburger2013realization, wili2023accordion, impertro2025strongly}.

The structure of this paper is as follows. We begin in Sec.~\ref{Sec:BH_model} by introducing the model and lattice geometry, as well as describing the relevant properties of the SF and RMI phases. We then discuss the transition between these phases in Sec.~\ref{sec:results}, including the extrapolation of the critical phase boundary to the thermodynamic limit. Sec.~\ref{sec:variances} describes how the different phases can be accessed experimentally, with a particular focus on how  the population and parity variances can be used to characterize the system. In Sec.~\ref{sec:geometries} we examine a wider set of ladder-like lattices, showing how the RMI phase generalizes for different quasi-1D geometries. Finally, Sec.~\ref{sec:end} presents our conclusions and outlines potential directions for future work.

\begin{figure}
    \centering
    \includegraphics[width=0.99\linewidth]{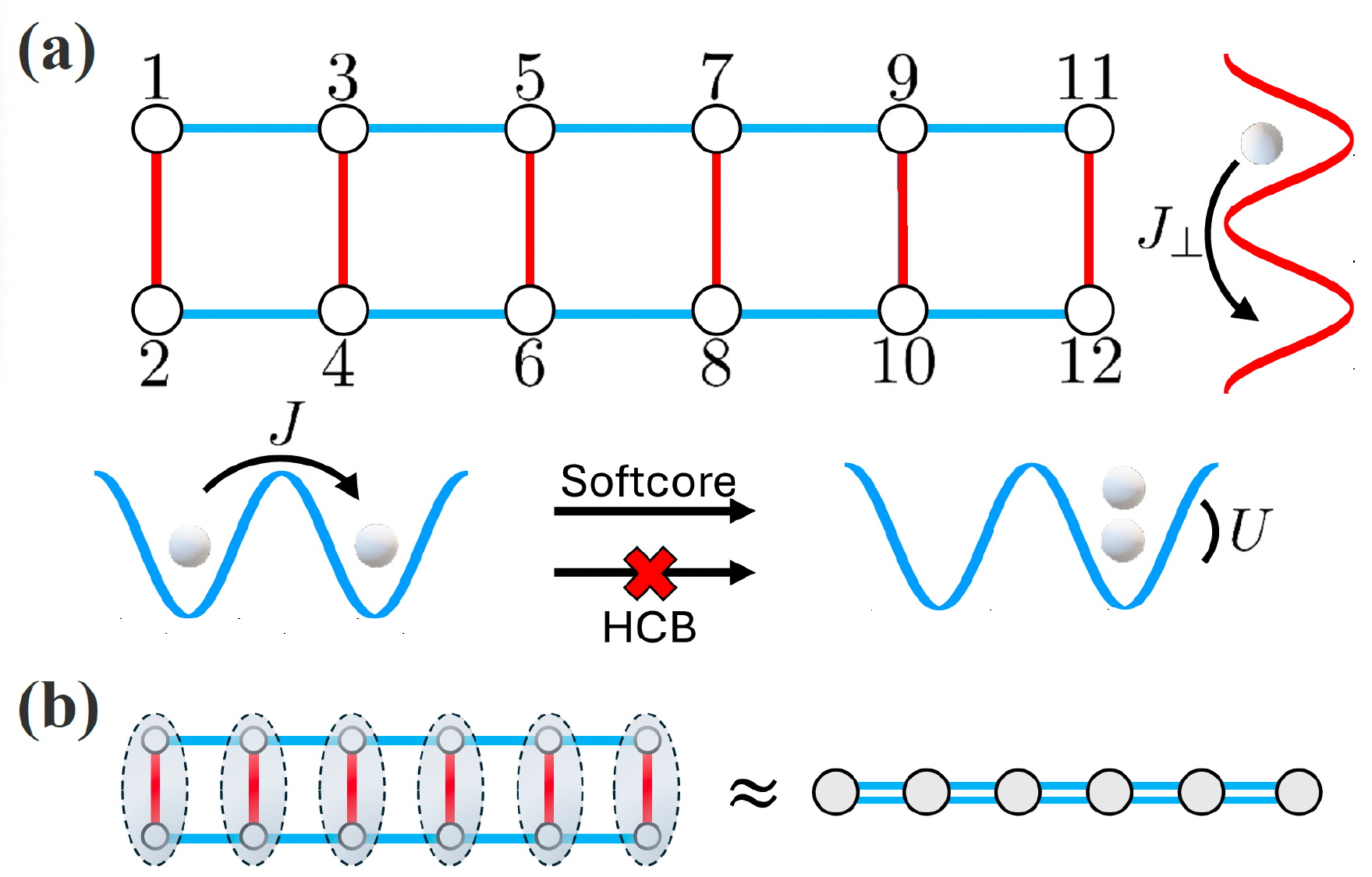}
    \caption{(a) Geometry of the two-leg ladder. Hopping along each chain occurs at rate $J$  (blue links), while the inter-chain hopping across the rungs at rate $J_\perp$ (red links). In the HCB limit ($U\rightarrow\infty$) doubly occupied sites are prohibited. (b) In the RMI phase each boson is delocalized across the two sites of a rung but localized along the chains, leading to insulating behavior. This configuration is analogous to the MI phase of a fully filled single chain with an effective hopping amplitude $2J$.}
    \label{fig:ladder_geometry}
\end{figure}

\begin{figure*}
    \centering
    \includegraphics[scale=0.49]{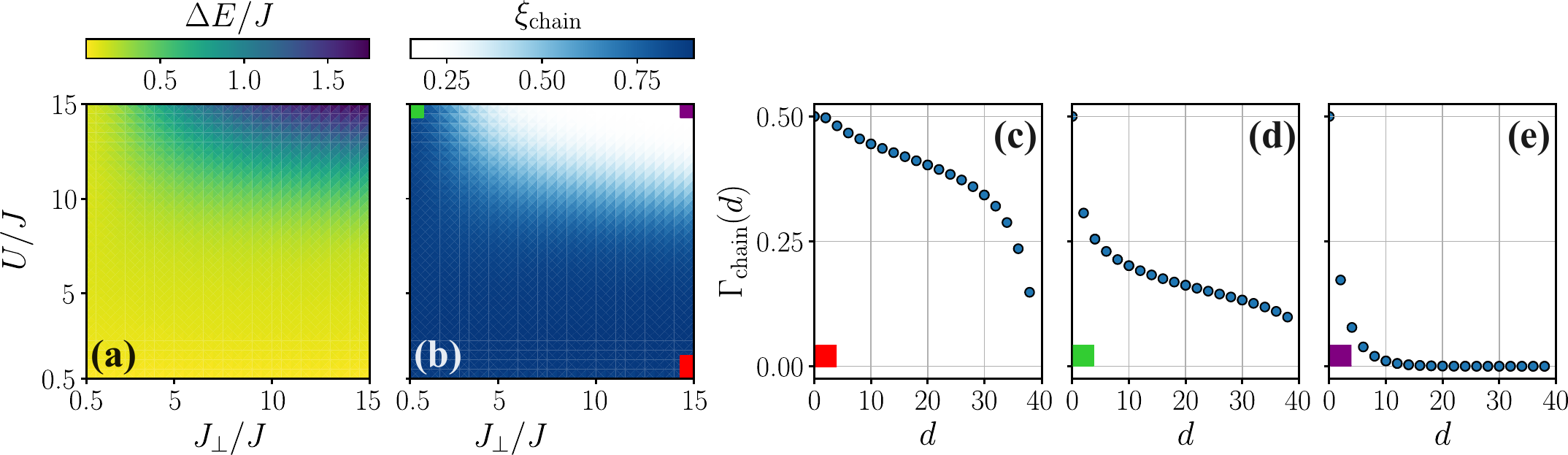}
    \caption{
    Phase diagrams showing (a) the energy gap, $\Delta E$, and (b) the normalized correlation length, $\xi_\text{chain}$, along the chains for a ladder with $\rho=0.5$ and $L=40$. The hopping correlation function, $\Gamma_\text{chain}$, for (c) $U/J=1$ and $J_\perp/J=15$, (d) $U/J=15$ and $J_\perp/J=0.5$, (e) $U/J=J_\perp/J=15$. The colored markers show the location of these parameters in the phase diargam of panel (b).}
    \label{fig:phase_diagrams}
\end{figure*}

\section{\label{sec:model}Bose-Hubbard Model on a ladder}\label{Sec:BH_model}

We begin by introducing the model describing the behavior of bosons on a ladder lattice. The system is governed by the Bose-Hubbard (BH)~\cite{bloch2008many,krutitsky2016ultracold, jaksch1998cold, kuhner1998phases} Hamiltonian,
\begin{equation}\label{eq:BH_general}
\hat{H} = \frac{U}{2}\sum_{i}^M\hat{n}_{i}(\hat{n}_{i}-1) - \sum_{i,j}^M J
_{ij} (\hat{a}^\dag_{i}\hat{a}_{j} + \hat{a}^\dag_{j}\hat{a}_{i}),
\end{equation}
where $U$ is the on-site interaction parameter and $J_{ij}
$ is the hopping rate from site $i$ to site $j$, defined by the lattice geometry through the hopping matrix $\mathbf{J}$. 
The sum in the second term runs over all pairs of sites which are connected by the lattice geometry. The Hamiltonian contains the creation (annihilation) operators $\hat{a}_i^\dag$ ($\hat{a}_i$) and the number operator $\hat{n}_i=\hat{a}_i^\dag\hat{a}_i$ at site $i$. Eq.~\eqref{eq:BH_general} describes a system with a fixed number, $N$, of bosons distributed over $M$ sites, giving a filling factor $\rho=N/M$.

For the ladder lattice geometry considered in this work, $\mathbf{J}$ is characterized by two parameters, $J$ and $J_\perp$, which connect neighboring sites [Fig.~\ref{fig:ladder_geometry}(a)].
The intra-chain hopping rate, $J$, describes tunneling between neighboring sites along the same leg of the ladder, while 
$J_\perp$ denotes the inter-chain (rung) hopping rate, connecting sites on opposite legs of a rung.
The RMI phase has been predicted for a half-filled ($\rho=1/2$) ladder of HCBs, and is characterized by each boson being delocalized over the two sites making up a rung. With the hardcore constraint, the RMI emerges as soon as $J_\perp\neq0$, but becomes more easily visible when $J_\perp\geq J$~\cite{crepin2011phase}. 

Another way the RMI can be described is by imagining each rung making up a single site of a 1D BH chain. As $J_\perp$ increases, the sites making up the rungs become more and more correlated, delocalising the population over the rungs, which can then be treated as the single site of a 1D chain [Fig.~\ref{fig:ladder_geometry}(b)]. 
This only works for $\rho\leq1/2$, as higher densities imply that bosons share the same rung, becoming impossible to map on a 1D chain in the HCB limit, which allows only one boson per site.. This picture is useful in demonstrating the influence of population density and lattice geometry on insulating phases, suggesting that similar MI phases appear in different quasi-1D ladder lattice geometries.

\subsection{The SF-to-RMI transition for softcore bosons}\label{sec:results}

The full phase diagrams for the ladder lattice at $\rho=1/2$ and $L=40$ were calculated using the density matrix renormalization group (DMRG)~\cite{schollwock2005density, schollwock2011density, orus2019tensor} algorithm as implemented in the \textsc{ITensor} library~\cite{Itensor1, Itensor2}.
The matrix product operators were constructed by labeling the lattice sites sequentially along the rungs  [Fig.~\ref{fig:ladder_geometry}(a)]. 
The DMRG calculation performs 150 sweeps to obtain the ground state, and 250 sweeps for the first excited state. Singular values smaller than $10^{-8}$ were discarded, and the maximum occupation number per site was limited to 6. This routine returns the ground state and first excited state eigenenergies, $E_{\text{GS}}$ and $E_{\text{e1}}$, and the single-particle density matrix, $\mathbf{p}$, with elements $p_{ij}=\langle \hat{a}^\dag_i\hat{a}_j\rangle$, which allows us to calculate all observables of interest. For simplicity, we use $J=1$ as our unit of energy, so that the parameters of the model are dimensionless. 

The opening of the energy gap, $\Delta E = E_{\text{e1}} - E_{\text{GS}}$, occurs for $U/J\gtrsim 10$ and $J_\perp/J\gtrsim 1$ signifying the presence of the RMI phase [Fig.~\ref{fig:phase_diagrams}(a)]. As $U$ is reduced, the BH model enters the SF phase~\cite{kuhner1998phases}. Hopping along the chains of the ladder is favored if $J_\perp$ is not strong enough, resulting in delocalization and the emergence of a SF phase. 

Although experimentally complex to measure, the mass gap helps visualize when the system enters an insulating phase. To support the observation of a SF-to-RMI transition, we evaluate the hopping correlation function as the average of all of the $p_{ij}$ terms which share the same separation $d$,
\begin{equation}\label{eq:gamma}
    \Gamma(d)=\frac{\sum_{|\mathbf{x}_i-\mathbf{x}_j |=da} p_{ij}}{\sum_{|\mathbf{x}_i-\mathbf{x}_j |=da} 1 },
\end{equation} 
where $d$ is expressed in units of lattice spacing $a$. We introduce the correlation functions $\Gamma_{\text{rung}}(d)$ and $\Gamma_{\text{chain}}(d)$, which are defined in the same way as $\Gamma(d)$, but include only terms which link the sites within a rung and a chain, respectively. 

For $U/J=1$ and $J_\perp/J=15$, we find that $\Gamma_{\text{chain}}(d)$ decays slowly following a power law, with a rapid decay at large $d$ [Fig.~\ref{fig:phase_diagrams}(c)] due to the open boundary conditions~\cite{lewenstein2012ultracold}. The strong correlations and noticeable influence of the boundary conditions signal the system is in a SF phase. This still holds true for $U/J=15$ and $J_\perp/J=0.5$, although $\Gamma_\text{chain}(d)$ decays faster than in the previous case [Fig.~\ref{fig:phase_diagrams}(d)]. On the other hand, if $U/J=J_\perp/J=15$, then $\Gamma_{\text{chain}}(d)$ decays exponentially [Fig.~\ref{fig:phase_diagrams}(e)]. This change from a power-law to an exponential decay is evidence of a BKT-type phase transition into an insulating phase~\cite{lewenstein2012ultracold}.


As shown in Tab.~\ref{tab:g_rung}, the sets of parameters corresponding to Figs.~\ref{fig:phase_diagrams}(c) and (e)  show a constant $\Gamma_{\text{rung}}(d)$, meaning the two chains of the ladder are strongly correlated with each other. A consequence of the exponential decay of $\Gamma_{\text{chain}}(d)$, while $\Gamma_{\text{rung}}(d)$, is constant is that each particle is strongly localized to a single rung while at the same time delocalized over its two sites. This is the hallmark of the RMI phase which can be also observed in the population density distribution [Fig.~\ref{fig:pop}(a)], where each boson occupies exactly one rung.
Tab.~\ref{tab:g_rung} also shows $\Gamma_{\text{rung}}(1)$ decreasing in the $J_\perp\rightarrow0$ limit, as in this case the chains are only very weakly coupled. 

\begin{table}
    \centering
\begin{tabular}{ c  c ||c |c | c c}
    &  & $U/J$ & $J_\perp/J$ & $\Gamma_\mathrm{rung}(0)$ & $\Gamma_\mathrm{rung}(1)$ \\ \hline
 (c) & \textcolor{fig_red}{$\blacksquare$} & 1 & 15  & 0.500 & 0.500\\ 
 (d) & \textcolor{fig_green}{$\blacksquare$}  & 15 & 0.5 & 0.500 & 0.338\\  
 (e) & \textcolor{fig_purple}{$\blacksquare$}  & 15 & 15 & 0.500 &0.498\\   
\end{tabular}
    \caption{Hopping correlation function $\Gamma_{\text{rung}}(d)$ along the rungs. The colored markers are used to refer each set of parameters to the points and graphs highlighted in Fig.~\ref{fig:phase_diagrams}.}
    \label{tab:g_rung}
\end{table}

The different phases can also be distinguished by the correlation length of the system, which is defined as~\cite{kuhner1998phases},
\begin{equation}\label{eq:corr_length}
    \xi^2 = \left(\frac{\sum_{d=0} 1}{\sum_{d=0} d^2}\right)\frac{\sum_{d=0} d^2\Gamma(d)}{\sum_{d=0} \Gamma(d)}.
\end{equation}
From this expression we can extract two correlation lengths, $\xi_{\text{rung}}$ and $\xi_{\text{chain}}$, measuring the decay of correlations along the rungs and chains, respectively. These correlation lengths are also normalised so to lie between 0 and 1. The maximum possible correlation length corresponds to a pure SF, where each site is maximally correlated to every other site, and $\Gamma_\text{chain}(d)$ and $\Gamma_\text{rung}(d)$  are constant. We observe $\xi_\text{chain}$ decrease towards 0 as $\Delta E$ increases [Fig.~\ref{fig:phase_diagrams}(b)]. 


The behavior we observed for finite-sized systems is indicative of the existence of a SF-to-RMI BKT-type phase transition for softcore bosons~\cite{kenna1997kosterlitz}. We calculate the location of the transition in the thermodynamic limit via scaling analysis~\cite{pai2005superfluid, luthra2008phase}. More details on the scaling procedure can be found in App.~\ref{app:Scaling}. 

\begin{figure}
\centering\includegraphics[width=0.99\linewidth]{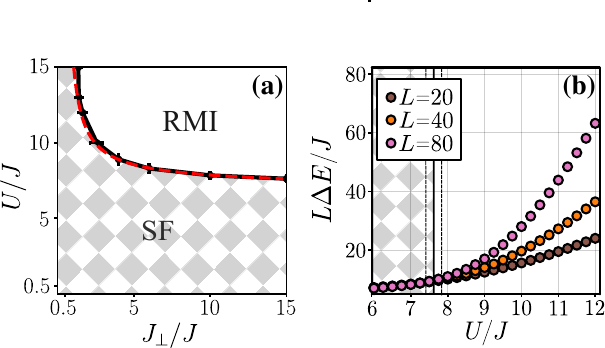}
    \caption{(a) Phase diagram in the thermodynamic limit for the half-filled ladder lattice. The RMI (white) and SF (checked) phases are highlighted. The phase boundary (solid, black line) includes errorbars showing thedeviation from results obtained with a more lenient cutoff [App.~\ref{app:Scaling}]. The approximate boundary from Eq.~\eqref{eq:boundary_intext} is shown as a red dashed line. (b) Example scaling analysis for $J_\perp/J=15$. The critical point was derived by comparing the data for different values of $L$. The extracted value is marked with a black vertical line while the black dashed borders display the error in this estimate.}
    \label{fig:phase_transition_scaled}
\end{figure}

The numerical data shows the critical value of the on-site interaction $U_c/J$ converging to a finite fixed value when $J_\perp/J\rightarrow\infty$ .
An analytical approximation for large $J_\perp$ describing the SF-to-RMI critical boundary is given by~\cite{crepin2011phase, svistunov1996superfluid, schonmeier2014ground}
\begin{equation}\label{eq:boundary_intext}
    \frac{J_{\perp,c}}{J} \approx A\exp\left[\frac{\pi}{4B\sqrt{U_c/2U^{1\text{D}}_c-1}}\right],
\end{equation}
where $A$ and $B$ are fitting constants, and $U^{1\text{D}}_c=3.25J$ is the critical point for the BH model on a 1D chain at commensurate filling~\cite{ejima2011dynamic}. A brief discussion on how the results from Refs.~\cite{crepin2011phase, svistunov1996superfluid, schonmeier2014ground} were adapted to obtain this equation for our system is given in App.~\ref{app:boundary}. Fitting the extrapolated numerical data with Eq.~\eqref{eq:boundary_intext} gives $A=0.235$, $B=0.457$, showing good agreement across the whole phase diagram, even when $J_\perp$ approaches $J$ [Fig.~\ref{fig:phase_transition_scaled}(a)]. 

The expression in Eq.~\eqref{eq:boundary_intext} shows that the critical hopping rate, $J_{\perp,c}$, diverges when the on-site interaction approaches $2U_c^{1D}$, with no phase transition possible for lower on-site interactions. Hence, the critical on-site interaction for a ladder with infinitely strong $J_\perp$ is equal to $2U_c^{1D}$, which matches the intuition of the rungs acting as single sites on a 1D chain for strong enough $J_\perp$.

\subsection{Number and parity variances}\label{sec:variances}

We are able to distinguish between the different phases present in the model by using the on-site and rung number variances, $\kappa$ and $\kappa_\text{rung}$, defined as
\begin{align}\label{eq:number_variance1}
    \kappa &=\frac{1}{M}\sum_{l, i}^M\langle \hat{n}_{l,i}^2 \rangle - \langle \hat{n}_{l,i}\rangle^2 ,\\
\label{eq:number_variance2}
    \kappa_\text{rung} &= \frac{1}{L} \sum_{i}^L\langle \hat{n}_{\text{rung}, i}^2 \rangle - \langle \hat{n}_{\text{rung},i}\rangle^2 ,
\end{align}
where $\hat{n}_{l, i}$ is the number operator acting on the site of the the $i$-th rung and $l$-th chain, and 
$\hat{n}_{\text{rung}, i} =\hat{n}_{1, i} + \hat{n}_{2, i}$ is the number operator for said rung. The change in labeling convention is to simplify reading for the ladder case. The convention used in Fig.~\ref{fig:ladder_geometry} can be reobtained with $(l,i)\rightarrow j= [4i-1+(-1)^l]/2$.
These parameters can be interpreted as indicators of the population delocalization, with $\kappa_\text{rung}$ treating each rung as a single site and therefore only quantifying contributions from the correlations along the chains.

\begin{figure}
\includegraphics[scale=0.4]{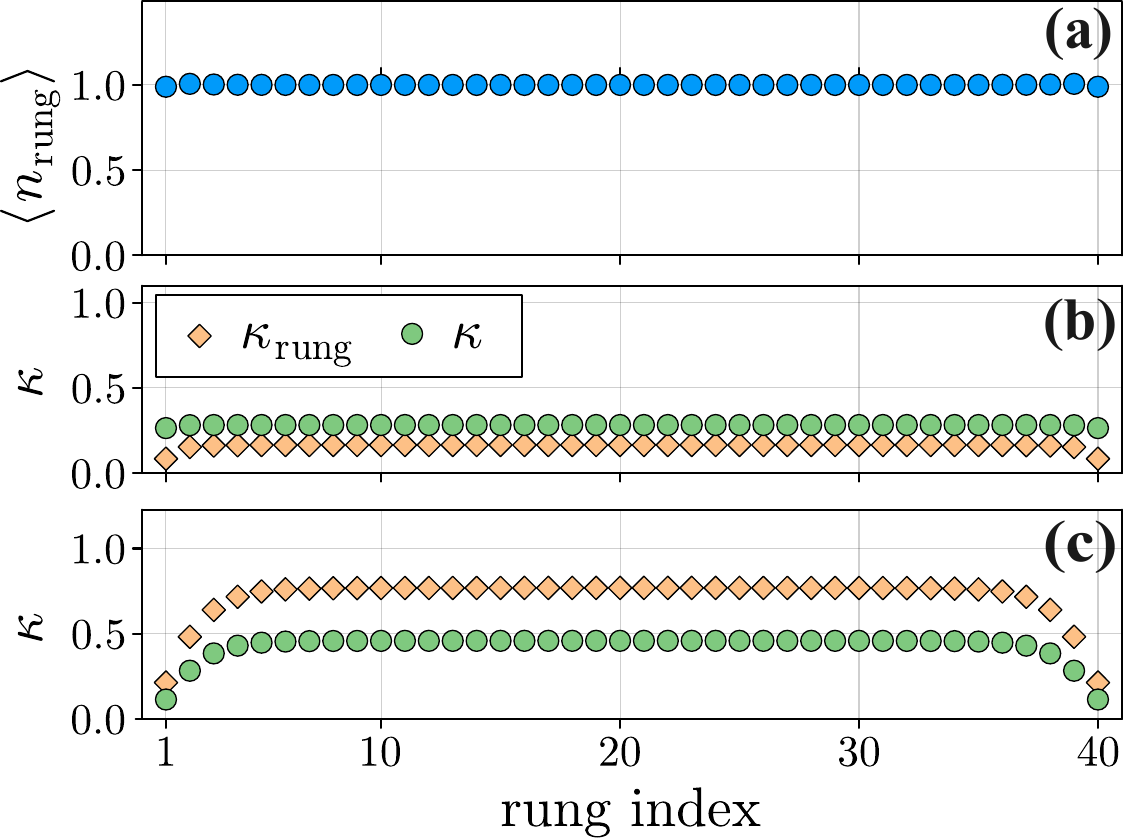}
\caption{(a) Population density $\langle n_\text{rung}\rangle$ and (b) corresponding population variances $\kappa$ and $\kappa_\text{rung}$ for $U/J=J_\perp/J=15$. The value of $\kappa$ is calculated using only the first chain. (c) Population variances for $U/J=1$ and $J_\perp/J=15$.}
\label{fig:pop}
\end{figure}

As the interaction strength is increased towards the HCB limit, $\kappa$ approaches $1/4$, as in this limit the on-site population is reduced to either 0 or 1, while $\kappa_\text{rung}$ approaches zero [Figs.~\ref{fig:pop}(b) and (c)]. For $J_\perp\rightarrow0$, the RMI phase has a slow opening of the gap such that $\Delta E/J<10^{-2}$ before growing substantially in the $1<J_\perp/J<2$ regime, even in the HCB limit~\cite{crepin2011phase}. This regime is characterised by $\langle\hat{n}_{1,i}\hat{n}_{2,i}\rangle\approx1/4$, leading to $\kappa_\text{rung}\approx1/2$.
If $\kappa_\text{rung}$ is large, so is the rung population variance, which means the bosons on each rung are  more likely to be delocalized along the chains, as expected for a SF phase. As $\kappa$ increases, it is still bounded between $1/4$ and $1/2$, while $\kappa_\text{rung}$ ranges from $0$ to $1$. Comparing the values of $\kappa$ and $\kappa_\text{rung}$ allows us to approximately determine if the system is in the SF phase, where $\kappa<\kappa_\text{rung}$, or the RMI phase, where $\kappa>\kappa_\text{rung}$.

Most quantum-gas microscope experiments only measure the parity of the atom-number on each site, as doubly-occupied sites are detected as empty due to light-assisted collisions. 
We now verify how such a parity projection influences the variances discussed above. We define the parity operator at each site~\cite{di2024commensurate}
\begin{equation}
    \hat{s}_{i, j}=\frac{1-(-1)^{\hat{n}_{i,j}}}{2},
\end{equation}
which is 0 if the population is even and 1 if it is odd. From this we can define the rung parity operator, 
\begin{equation}
\hat{s}_{\text{rung},j}=\hat{s}_{1,j}+\hat{s}_{2,j}-2\hat{s}_{1,j} \hat{s}_{2,j}.
\end{equation} 
The on-site and rung parity variances are defined analogously to Eqs.~\eqref{eq:number_variance1} and~\eqref{eq:number_variance2} and using $\hat{s}_{i, j}$ and $\hat{s}_{\text{rung}, j}$ instead of $\hat{n}_{i, j}$ and $\hat{n}_{\text{rung}, j}$. The average parity variances are defined as
\begin{align}
    \sigma =S(1-S), && \sigma_\text{rung}=S_\text{rung}(1-S_\text{rung}),
\end{align}
where $S$ and $S_\text{rung}$ are the average on-site and rung parities,
\begin{align}
    S=\frac{1}{M}\sum_{l,j}^M\langle\hat{s}_{l,j}\rangle,&&S_\text{rung}=\frac{1}{L}\sum^L_j\langle\hat{s}_{\text{rung},j}\rangle.
\end{align}

\begin{figure}
\centering    \includegraphics[width=1.0\linewidth]{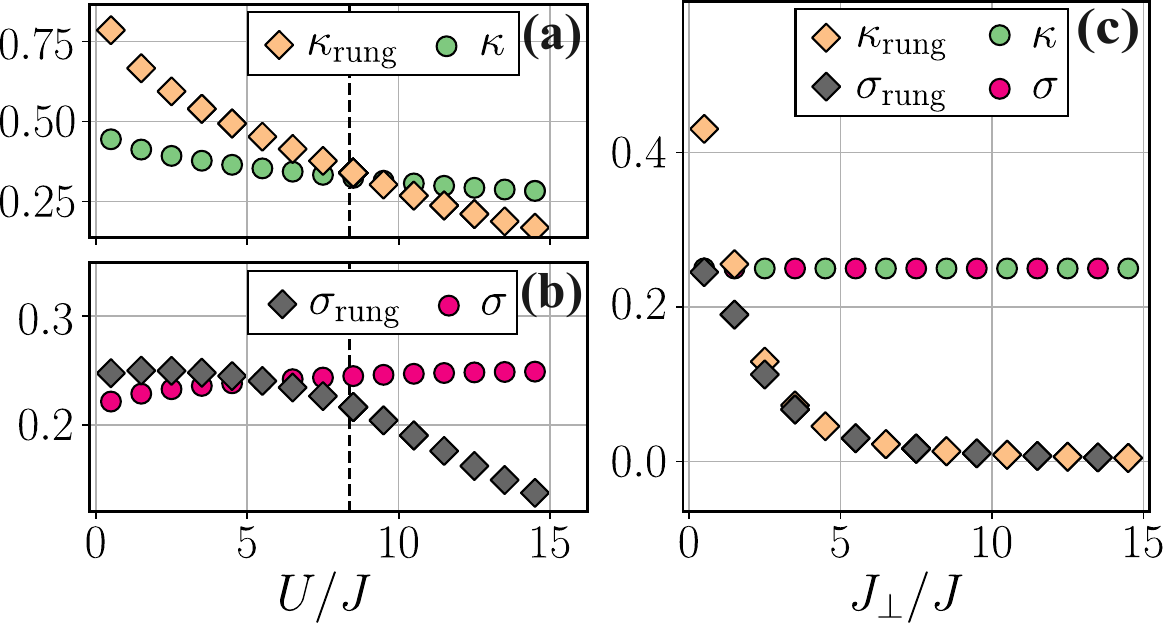}
    \caption{Average (a) on-site and (b) rung number variances, $\kappa$ and $\kappa_\text{rung}$, along with the corresponding parity variances, $\sigma$ and $\sigma_\text{rung}$, calculated for a half-filled ladder ($L=40$) at $J_\perp/J=10$. The dashed black line is located at the critical point, calculated via scaling analysis. (c) Same quantities in the HCB limit, where the on-site variances become identical, while the rung variances are distinguishable for $J_\perp\lesssim 4J$.}
\label{fig:variance_compare}
\end{figure}
While for most of the phase diagram $\kappa$ and $\sigma$ show similar behavior, differences arise for small $U$. As $U$ increases, the on-site parity variance, $\sigma$, increases towards $1/4$, while $\kappa$ decreases towards it instead [Figs.~\ref{fig:variance_compare}(a) and (b)]. This is in line with the average on-site parity being 1/2 in the RMI phase due to the half-filling condition.
On the other hand, $\sigma_\text{rung}$ decays together with $\kappa_\text{rung}$, mirroring the behavior of population variances for a 1D chain. The variation of $\sigma_\text{rung}$ across the phase diagram  provides an experimentally accessible probe of the different physical behavior  in the SF and RMI phases.
The differences between $\kappa$ and $\kappa_\text{rung}$ when compared to $\sigma$ and $\sigma_\text{rung}$ arise from the contributions of states with highly populated sites. These tend to increase $\kappa$ and decrease $\sigma$ because of the inclusion of additional states with even parity.

Since sites with double and higher occupancy are forbidden in the HCB limit, $\kappa$ and $\sigma$ are identical and independent of $J_\perp$ [Fig.~\ref{fig:variance_compare}(c)]. Differences emerge when considering the rung-based quantities instead, as both converge and decay towards zero as $J_\perp$ increases but are distinguishable for $J_\perp\rightarrow0$. The difference is explicitly calculated as
\begin{equation}
    \kappa_\text{rung} - \sigma_\text{rung} = \left(\frac{2}{L}\sum_{i=1}^L\langle \hat{n
    }_{1,i}\hat{n}_{2,i}\rangle\right)^2.
\end{equation}
The difference between $\kappa_\text{rung}$ and $\sigma_\text{rung}$ is related to the average density correlation along the rungs. Hence, an increase in this difference measures an increase of states with doubly occupied rungs.
Bosons are then more prone to delocalize over the rungs for small $J_\perp$, although they would not be able to share sites due to the HCB condition.

\section{Generalized quasi-1D geometries}\label{sec:geometries}

\begin{figure}[t!]
    \includegraphics[width=0.85\linewidth]{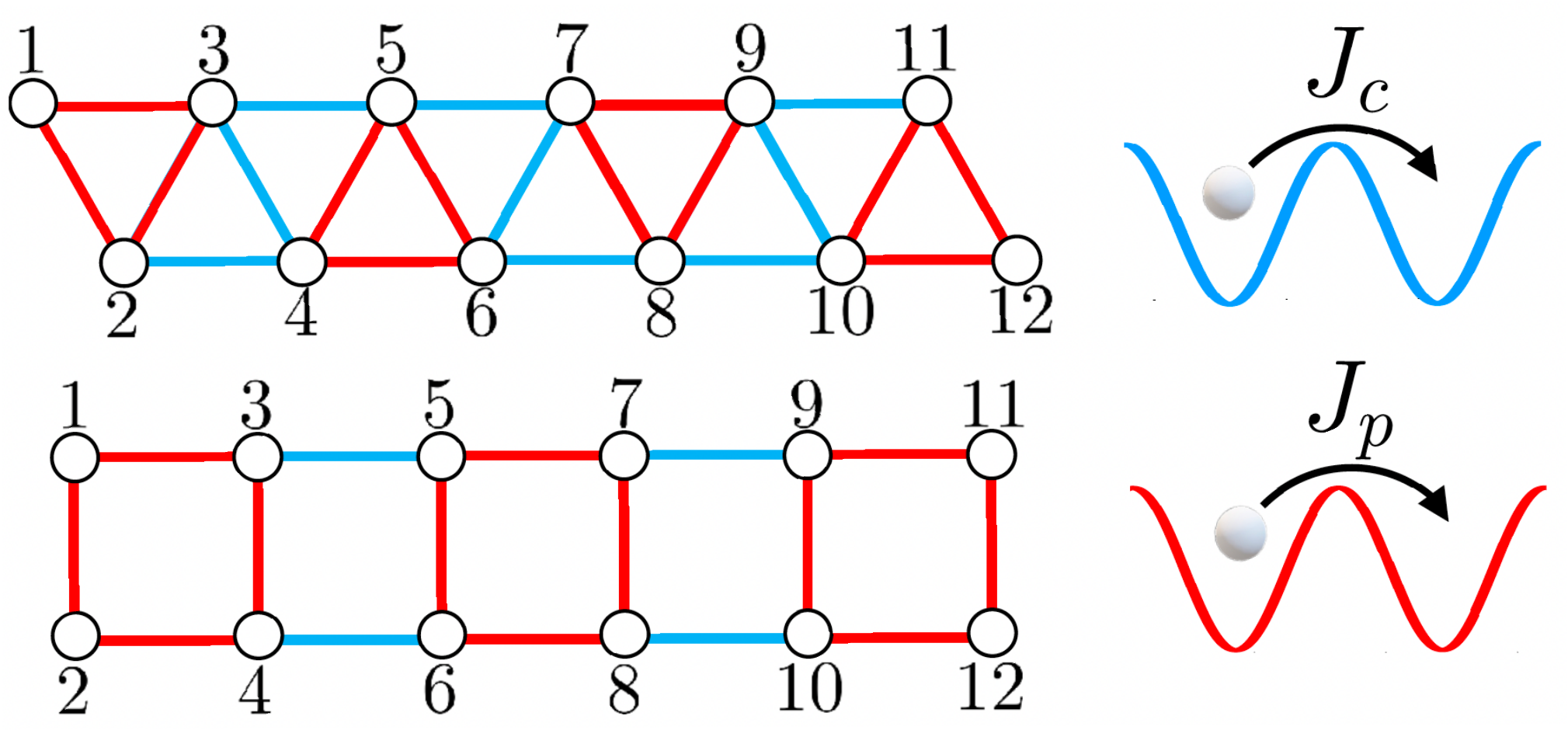}
    \caption{Lattice geometries of the triangular ladder (top) and square ladder (bottom). The $J_c$ ($J_p$) hopping rate couples the sites connected by blue (red) lines.}
    \label{fig:geometries}
\end{figure}

The phases observed in the ladder lattice described above include the insulating RMI phase, which requires a density of one particle per rung. This can be attributed to the geometrical properties of the model. Next we investigate whether similar phases arise in two other quasi-1D geometries comparable to the ladder lattice. 
These are the triangular ladder, which is made up of a chain of triangular plaquettes with population density, $\rho=1/3$, and the square ladder, also referred to as the uniform SSH ladder~\cite{padhan2024interacting} and made up of a chain of square plaquettes with population density, $\rho=1/4$.
For these the hopping matrix, $\mathbf{J}$, is described by two parameters. These are $J_{c}$, the hopping rate between sites in different plaquettes, and  $J_{p}$, the hopping rate between sites within the same plaquette [Fig.~\ref{fig:geometries}].
We expect the influence of finite on-site interactions on these more complex geometries to be analogous to that of the standard ladder lattice. Thus, we limit our analysis to the HCB case.

To identify the presence of an insulating phase, we again analyze the energy gap, $\Delta E$, and the correlation length along the chains, $\xi_\text{chain}$. 
We define the correlation length within a single plaquette, $\xi_\text{plaq}$, as obtained from the plaquette correlation function, $\Gamma_\text{plaq}(d)$, which includes only correlations between sites belonging to the same plaquette. These quantities are calculated in an analogous way to Eqs.~\eqref{eq:gamma} and \eqref{eq:corr_length}. The plaquette correlation length, $\xi_\text{plaq}$, is also normalized as described in Sec.~\ref{sec:results}.

\begin{figure}
    \centering\includegraphics[width=0.775\linewidth]{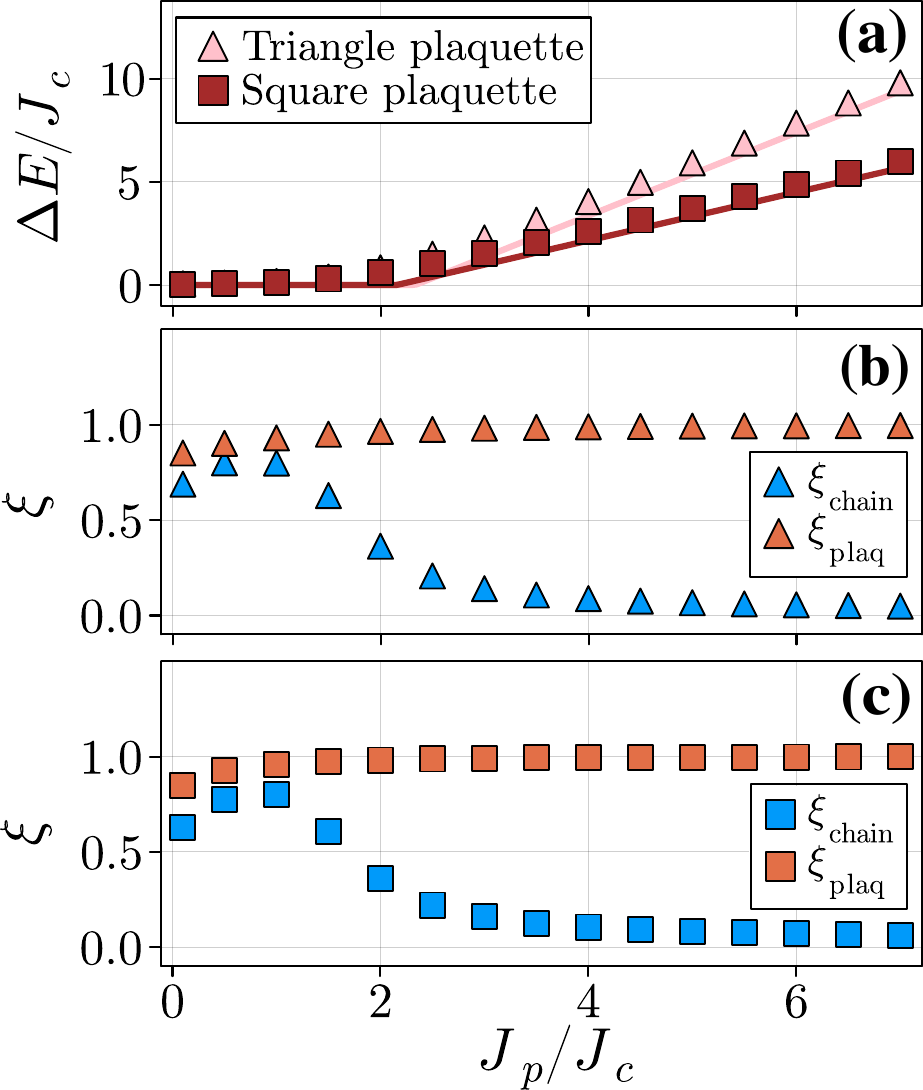}
    \caption{Phase transition in the triangular ($\rho=1/3$) and square ($\rho=1/4$) plaquette ladders, both of size $L=42$.  (a) The energy gap, $\Delta E /J_c$,. The numerical results (symbols) are compared to analytic result (solid lines) described in the main text. The normalized  correlation lengths, $\xi_{\text{chain}}$ and $\xi_{\text{plaq}}$ are shown for (b) the triangular plaquette ladder and (c) the square plaquette ladder.}
    \label{fig:triangle_ssh_HCB}
\end{figure}

The relevant BH Hamiltonian was solved via DMRG using the same convergence parameters as in the previous section. We find that the energy gap, $\Delta E$, opens in a similar way to the standard ladder geometry and is linear in $J_p$ when $J_p\gg J_c$ [Fig.~\ref{fig:triangle_ssh_HCB}(a)]. 
The gap of the square plaquette ladder opens more slowly than in the triangular ladder. This is contrary to naive expectations from the ladder geometry. The triangular plaquettes are more densely connected, with more hopping paths between neighboring plaquettes than the square ladder. This increased number of connections could make the opening of the energy gap less favorable. However, the triangular plaquettes are also completely connected, such that a boson can directly hop between all the sites making up a plaquette. This effect dominates, and makes the gap in the triangular lattice open more quickly than in the square ladder. 

By calculating $\Delta E$ perturbatively for $J_c\ll J_p$, the energy gap of ladders of completely connected plaquettes, such as the triangle ladder, approaches the free-fermion limit [App.~\ref{sec:lin_lim}],
\begin{equation}\label{eq:linear_lim}
    \Delta E = 2J_p-2\left[\frac{3m-2}{m^2} \right]CJ_c + \mathcal{O}(J_c^2).
\end{equation}
Here, $m$ is the number of sites per plaquette and $C$ the number of connections between a pair of plaquettes. This expansion gets more precise deeper in the insulating regime, where $J_p/J_c\gg1$. In this limit, completely connected plaquettes map onto single sites of a 1D chain. To zeroth order, the energy gap opens as $2J_p$, independently of the system size.

Both the triangular and square plaquette ladders can be described as ladders of rings, where the plaquettes have only nearest neighbor connections. The energy gap in such a system obeys [App.~\ref{sec:lin_lim}]
\begin{equation}\label{eq:linear_lim2}
    \Delta E = 4\left[1-\cos\left(\frac{\pi}{m}\right)\right]J_p-2\left[\frac{3m-2}{m^2} \right]CJ_c + \mathcal{O}(J_c^2).
\end{equation}
Plaquettes that are not fully connected, such as in the square ladder, show a weaker influence of $J_p$ and a slower opening of the energy gap. For ladders of plaquettes with equal $m$, Eq.~(\ref{eq:linear_lim2}) provides a lower bound for the energy gap. The dependence of $\Delta E$ on $J_p$ becomes weaker as plaquette size increases, progressively reducing its contribution. The first-order term of the expansion does not rely on the complete connection condition, and is thus more general. This term depends linearly on $C$ but decreases with increasing $m$. Since it only depends on $J_c$, it functions as the offset of $\Delta E/J_c$, determining when the gap opens.

The opening of the gap is accompanied by a decay of $\xi_{\text{chain}}$ for both geometries [Figs.~\ref{fig:triangle_ssh_HCB}(b) and (c)]. These features are especially evident around $J_p=J_c$, where the gap opens more significantly. The correlation length, $\xi_{\text{chain}}$, increases slightly at small $J_{p}$, as hopping along the chains is partly controlled by $J_p$ in both models. Nonetheless, both $\Delta E$ and $\xi_{\text{chain}}$ show the emergence of an insulating phase.
On the other hand, $\xi_{\text{plaq}}$ does not decay with increasing $J_{p}$, instead it quickly saturates to its maximum value for both geometries. This means that, just like in the standard ladder geometry, the MI phase consists of a single boson localized on each plaquette but delocalized over the sites making up that plaquette. For the chosen densities, this results in there being exactly one boson per plaquette. This is backed up by previous studies for the SSH ladder~\cite{padhan2024interacting}, where an insulating phase was also defined in a similar way. Thus, a plaquette-Mott insulator is observed for these generalized quasi-1D ladder geometries. The RMI is the special case of this for the standard ladder lattice. 

Here, we have only examined the case where the density is $\rho=1/m$. However, equivalent insulating phases also emerge at different densities. The triangular ladder shows such phases for $\rho=2/3$ due to particle-hole symmetry, and the square ladder reverts to the RMI phase for $\rho=1/2$~\cite{padhan2024interacting}. Thus, the plaquette-Mott insulator can always be observed when the total number of bosons is an integer multiple of $m$.

\section{Conclusions}\label{sec:end}

In this work, we investigated the emergence of the RMI phase in a half-filled bosonic ladder with finite on-site interactions, examined the phase diagram and characterized the SF-to-RMI transition. We calculated observables such as the on-site and rung variances, which can be directly measured with a quantum-gas microscope~\cite{gross2021quantum}.
Generalizing the ladder structure showed that a plaquette-Mott insulator can be found in quasi-1D lattices. This insulating phase arises for fractional fillings due to their commensurability with the plaquette size.
The lattice structures required for both the triangular~\cite{becker2010ultracold} and square~\cite{dong2025observation} plaquettes studied here have already been realized experimentally. Similarly, additional insulating phases have been predicted for multileg ladders, where Mott-like phases persist at fillings corresponding to one particle per rung~\cite{mishmash2011bose, block2011exotic, Potapova2023multileg}. Although increasing the number of legs reduces the energy gap, RMI-like phases should be observable in ladders with a few coupled chains.

Future studies may include a more detailed investigation of commensurability effects at different fillings, as well as the influence of disorder in different lattice structures and densities, which would provide further insight into geometry-induced localization.
In addition, ladder lattices continue to serve as a valuable platform to investigate topological effects~\cite{wienand2024emergence, wellnitz2025emergent}. This leads to the interesting possibility of using them to study both dynamics-induced~\cite{wintersperger2020realization} and interaction-induced~\cite{parida2024interaction, padhan2024interacting} topological physics.
Quasi-1D geometries can also provide insights into the physics of mobile impurities interacting with a bath~\cite{Lamacraft2009dispersion, Schecter2012Critical}. The ladder lattice has already been used to theoretically investigate the influence of dimensionality on the Anderson orthogonality catastrophe \cite{Kamar2025impurity}. Generalizing these ideas to more involved quasi-1D geometries may lead to the emergence of novel dynamical features. 

Finally, the ladder has been useful for studying the crossover between different dimensionalities~\cite{revelle20161d, yao2023strongly, Potapova2023multileg, guo2024observation}. This characteristic was used to observe the emergence of 2D physics from a 1D structure. Studying these effects using fractal geometries, which have a fractional dimensionality~\cite{Manna2022fractal, Manna_2024, Dutkiewicz2026fractal}, opens the door to studying the stability of the RMI and related states to disorder arising both from the Hamiltonian and the lattice geometry.

\acknowledgments{ 
We thank Arthur La Rooij, Christopher Parsonage and Lennart Koehn for helpful discussions. This work was supported by the Engineering and Physical Sciences Research Council (EPSRC): LC through the Doctoral Training Partnership (grant no. EP/W524670/1), SK through the Programme Grant 'Quantum Advantage in Quantitative Quantum Simulation' (EP/Y01510X/1) and PK through grant No. EP/Z533713/1. 
}

\appendix

\section{Scaling analysis}\label{app:Scaling}

Here we discuss the scaling analysis approach we used to determine the location of the SF-to-RMI transition~\cite{pai2005superfluid, luthra2008phase}.
A useful indication of this transition is the change in behavior of the correlation length, from a power-law decay in the SF to exponential decay in an insulating phase. The SF-to-RMI phase transition belongs to the BKT universality class, and is therefore properly defined in the thermodynamic limit, where $M\rightarrow\infty$ and $\rho$ remains finite. 

We used the finite-size scaling analysis close to the critical region to calculate the energy gap, which follows the standard scaling function~\cite{Sachdev2004book}
     \begin{equation}
    \Delta E\approx\frac{1}{L}f\left(\frac{L}{\xi_{\text{chain}}}\right),
    \end{equation}
where $f(x)$ is constant when $x\ll1$. Consequently, different curves for different values of $L\Delta E/J$ converge in the SF phase and diverge in an insulating phase.

We computed curves for $L=20, 40, 80$ using DMRG with a cutoff of $10^{-8}$, and derived transition points by varying $U$ at fixed $J_\perp$ and vice-versa . Given the increasing sizes of the matrix product operators, the number of singular-value decomposition sweeps of both the ground and first excited states, $|\text{GS}\rangle$ and $|\text{e1}\rangle$ respectively, were increased with system size [Tab.~\ref{tab:sweeps_scaling}].  
We consider the curves as converged when the difference between $L\Delta E/J$ of different system sizes is less than $5\cdot10^{-3}$. The standard error was calculated following a similar procedure and by imposing a more lenient cutoff of $10^{-2}$ [See Fig.~\ref{fig:phase_transition_scaled}(a) for main results and  Fig.~\ref{fig:phase_transition_scaled}(b) for example of $L\Delta E/J$ versus $U$ and $J_\perp/J=15$].

\begin{figure}[t!]
    \centering
    \includegraphics[width=1.\linewidth]{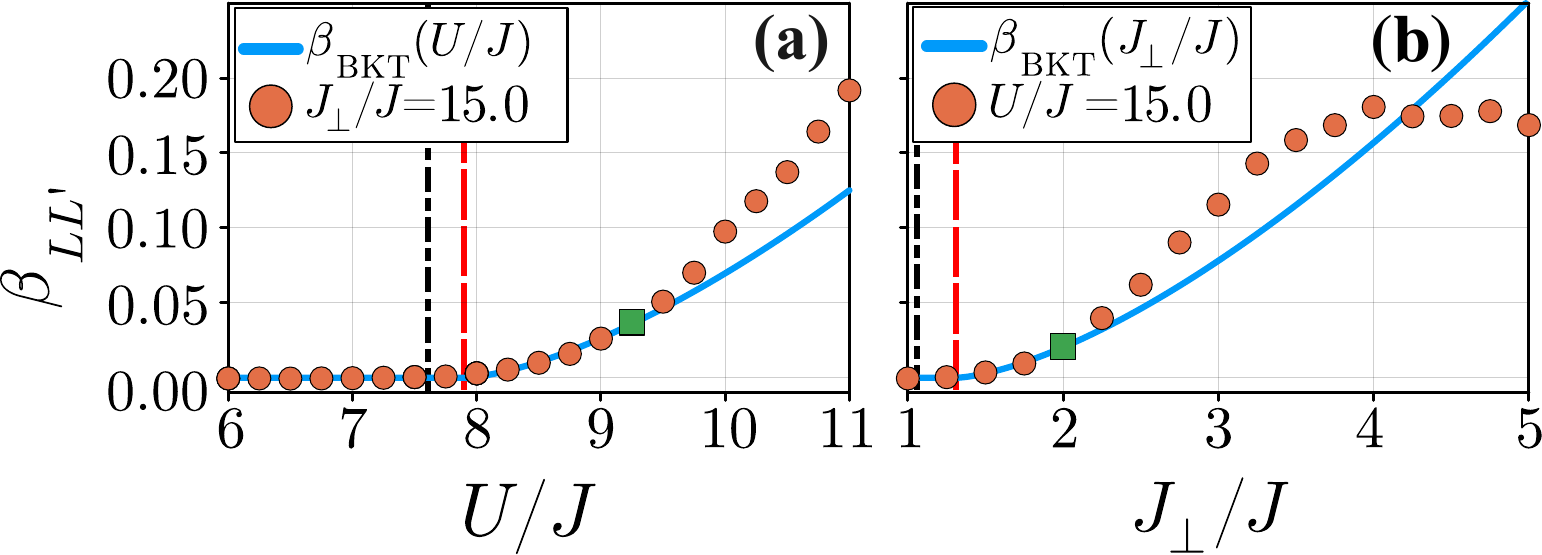}
    \caption{ (a) Numerical results of $\beta_{LL'}(g)$ [Eq.\eqref{eq:RW}] vs.~$U$ for fixed $J_\perp/J=15$, and (b) vs.~$J_\perp$ for fixed $U/J=15$. Blue lines are a fit with $\beta_\text{BKT}(g)$, and the green squares are the last points used in the fitting routine. The black dot-dashed line locates the critical point calculated using Eq.~\eqref{eq:boundary_intext}, while the red dashed one locates the critical point retrieved by fitting $\beta_\text{BKT}$ to the $\beta_{LL'}$ data.}
    \label{fig:RW_beta}
\end{figure}

\begin{table}
\centering
\begin{ruledtabular}
\begin{tabular}{lccc}
State & $L=20$ & $L=39,40$ & $L=80$ \\
\hline
$|\text{GS}\rangle$ & 60  & 150 & 180 \\
$|\text{e1}\rangle$ & 150 & 250 & 250 \\
\end{tabular}
\end{ruledtabular}
\caption{Number of sweeps imposed to converge with DMRG to the ground state $|\text{GS}\rangle$ and first excited state $|\text{e1}\rangle$ for different sizes $L$. The retrieved matrix product states were used for scaling analysis and calculation of the Roomany-Wyld approximant.}
\label{tab:sweeps_scaling}
\end{table}

To further confirm that the phase transition for the ladder is of BKT-type, we use the Roomany-Wylde approximant of the $\beta$-function~\cite{roomany1980finite, pai2005superfluid}
    \begin{equation}\label{eq:RW}
    \beta_{LL'}(g) = \frac{1 - \ln\left[\xi_{\text{chain}}(L)/\xi_{\text{chain}}(L')\right]/\ln(L/L')}{\sqrt{\xi_{\text{chain}}(L)\xi_{\text{chain}}(L') /\xi'_{\text{chain}}(L)\xi'_{\text{chain}}(L') }},
    \end{equation}
where $\xi'_{\text{chain}}(L) =\partial\xi_{\text{chain}}(L) /\partial g$, and $g$ is a parameter influencing the system, either $g=U$ or $g=J_{\perp}$. For Eq.~\eqref{eq:RW} to be valid, the chain lengths $L$ and $L'$ must be as close as possible, and we choose $L=40$, $L'=39$.

For a BKT-type transition, it has been shown that Eq.~\eqref{eq:RW} reduces to $\beta_{\text{BKT}}(g)=G(g-g_c)^{1+\gamma}$ close to the transition point at $g_c$, with the critical exponent $\gamma=0.5$~\cite{roomany1980finite} and $G$ as a free scaling constant. We verify the BKT nature of the phase transition by fitting $\beta_\text{BKT}$ to the calculated values of $\beta_{LL'}$ versus $U$ [Fig.~\ref{fig:RW_beta}(a)] and $J_\perp$ [Fig.~\ref{fig:RW_beta}(b)]. 

By performing this fitting, the critical point was found to be $U^{\text{RW}}_c/J=7.98$ for $J_\perp/J=15$, while Eq.\eqref{eq:boundary_intext} returns the critical value $U^a_c/J=7.61$, with a relative error between the two of $\delta_U=|U^{\text{RW}}_c-U^a_c|/J=0.37$. For $U/J=15$, we obtained the critical value via fitting to be $J^{\text{RW}}_{\perp, c}/J=1.31$, while Eq.\eqref{eq:boundary_intext} gives $J^a_{\perp, c}/J=1.06$, with a relative error between the two of $\delta_J=|J^{\text{RW}}_{\perp, c}-J^a_{\perp, c}|/J=0.25$.


\section{Calculating the phase boundaries}\label{app:boundary}

We here discuss how the work from Refs.~\cite{crepin2011phase, svistunov1996superfluid, schonmeier2014ground} was applied to obtain Eq.~\eqref{eq:boundary_intext}.
We start by considering the renormalization group (RG) flow equations from Eq.~(3.16) in Ref.~\cite{crepin2011phase} for correlated chains.
The equations were derived by applying bosonization, resulting in a Hamiltonian depending on the (anti-)symmetric Luttinger parameter, $K_{s(a)}$, and group velocity, $v$. The RG flow equations were obtained by defining four dimensionless coupling parameters: the anti-symmetric $y_a$, symmetric $y_s$, and inter-chain $y_\perp$ couplings, along with $g_s$, which is generated under RG flow. These coupling parameters are read as $y_\perp(a)=aJ_\perp/v$, $y_a(a)=y_s(a)=aJ_\perp v$, $g_s(a)=0$, with $a$ being the lattice periodicity.

In the limit of strongly correlated chains, we can reduce the RG flow equations such that they are only dependent on the relevant parameters $y_\perp$ and $K_a$:
\begin{align}
\frac{d\tilde{y}_\perp}{dl}&=\left( 2-\frac{1}{\tilde{K}_a}\right)\tilde{y}_\perp, \\
\frac{d\tilde{K}_a}{dl}&=(\tilde{y}_\perp)^2,
\end{align}
where we renormalised the parameters, $\tilde{y}_\perp=y_\perp/\sqrt{2}$ and $\tilde{K}_a=2K_a$. The flow parameter is defined as $l=\ln{\tilde{a}(l)}$, with $\tilde{a}(l)=a(l)/a$ being the unitless characteristic length scale $a(l)$. The equations are of the Sine-Gordon type, describing a set of rungs with vanishing $J/J_\perp$.

Following previous works~\cite{svistunov1996superfluid, schonmeier2014ground}, the critical boundary is found to behave as
\begin{equation}\label{eq:step1}
    \frac{(J_\perp)_c}{J} \propto\exp\left[\frac{\pi s}{4b\sqrt{U/U_0-1}}\right]
\end{equation}
for $J_\perp\gg J$, where the parameter $U_0$ is the critical transition point for the case of $J/J_\perp\rightarrow0$ and $s$ and $b$ are scaling constants.
Furthermore, we notice the value $s$ can be absorbed by the constant $b$ as $B=b/s$. 

\begin{figure}[t!]
    \centering
    \includegraphics[width=0.8\linewidth]{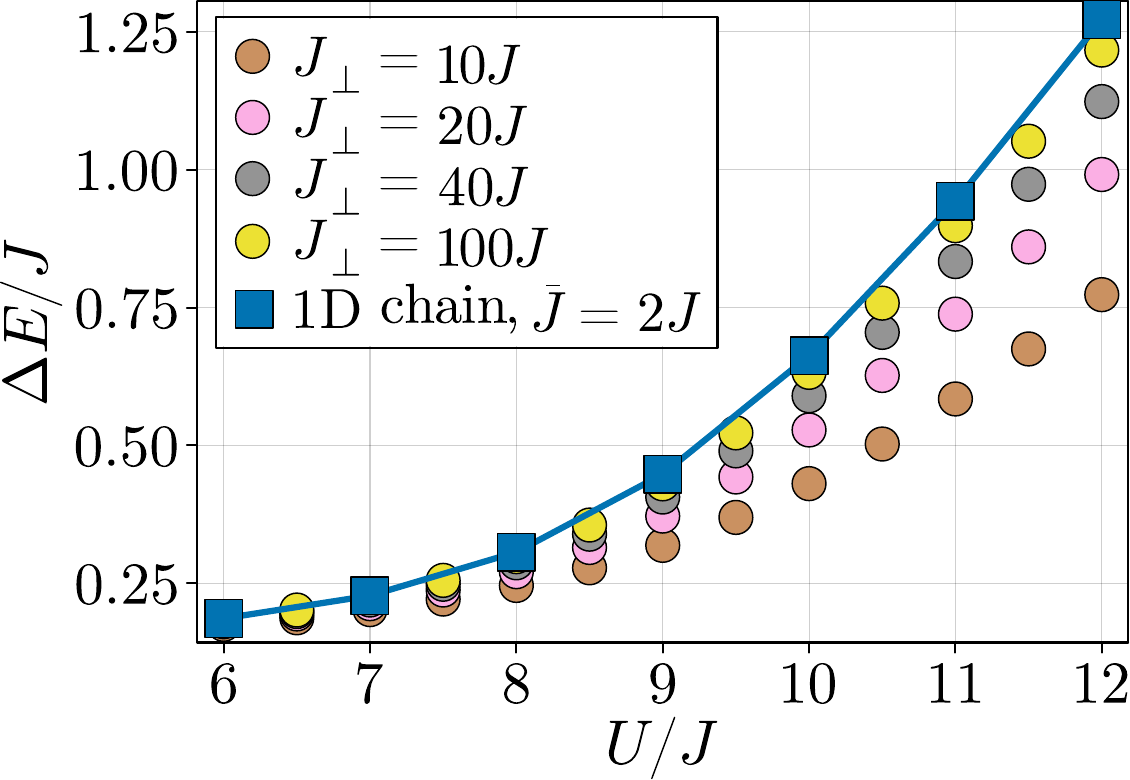}
    \caption{Energy gap of a half-filled ladder lattice vs.~interaction strength for different values of $J_\perp$ (circles), compared  to a 1D chain with an effective hopping parameter of 
$2J$ (squares). The line is to guide the eye.}
    \label{fig:jperp_lim}
\end{figure}

Since we impose $J=1$ for all our calculations, we define $U_0$ as the transition point as $J_\perp\rightarrow\infty$.
To determine $U_0$, we consider the ladder as a commensurate 1D chain. We expect $U_0=2U^{1\text{D}}_c$ due to the existence of two tunneling connections when looking at intrarung hoppings.
The idea is confirmed by observing the energy gap of a $\rho=0.5$ ladder converge to the behaviour of the gap of a $\rho=1$ chain with $2J$ hopping as $J_\perp\rightarrow\infty$ [Fig.~\ref{fig:jperp_lim}]. Thus, the SF-to-RMI boundary transition follows Eq.~\eqref{eq:boundary_intext}.

\section{Linear limit of the HCB mass gap for generalised ladders}\label{sec:lin_lim}

To exlain the derivation of the energy gaps for completely connected and ring plaquettes [Eqs.~\eqref{eq:linear_lim} and \eqref{eq:linear_lim2}], we
use perturbation theory~\cite{messiah2014quantum} and follow the approach of Ref.~\cite{crepin2011phase}. While our numerical simulations calculate $\Delta E = E_{\text{e1}}-E_{\text{GS}}$, Ref.~\cite{crepin2011phase} uses the charge gap, defined as $\Delta E = \epsilon(L+1)+\epsilon(L-1)-2\epsilon(L)$, where $\epsilon(N)=E_\text{GS}(N)$ is the ground state for $N$ bosons. However, for the SF-to-RMI transition, these quantities have the same behaviour in the thermodynamic limit.
We consider a chain of plaquettes, each made up of $m$ sites and $n$ bosons.
In the HCB limit, the Hamiltonian is $\hat{H}=-J_p\hat{H}_{\text{plaq}}-J_c\hat{H}_{\text{conn}}$, where $H_\text{plaq}$ defines the intra-plaquette connections and $H_\text{conn}$ the inter-plaquette ones. We define the energy of a single plaquette with $n$ bosons on $m$ sites as $E_{n,m}$.

\subsection{Completely connected plaquettes}

In the following, we discuss the derivation of the charge gap of ladders of completely connected plaquettes [Eq.~\eqref{eq:linear_lim}].
To calculate the zeroth order, we consider only a single plaquette. In this case, the $\hat{H}_{\text{plaq}}$ matrix spans a Hilbert space of dimension 
$d(n,m)=m!/[n!(m-n)!]$, with each row containing $n(m-n)$ nonzero elements, as each boson can hop to any unoccupied site.
Thus, the lowest eigenenergy is $E_{n,m} = -J_pn(m-n)$, with ground state $|n\rangle$.

The ground state of a chain of $L_p$ plaquettes and $L_p$ bosons is $|L_p\rangle=|1\rangle\otimes |1\rangle\dots \otimes |1\rangle$ with energy $\epsilon^{(0)}(L_p)=L_pE_{1,m}=-J_pL_p(m-1)$. If we consider a chain with $L_p+1$ bosons, one of the plaquettes must be doubly occupied, and the energy in this case is $\epsilon^{(0)}(L_p+1)=(L_p-1)E_{1,m}+E_{2,m}=-J_p(L_p-1)(m-1)-2J_p(m-2)$. Similarly, for the case of $L_p-1$ bosons, one plaquette must be empty, leading to $\epsilon^{(0)}(L_p-1)=(L_p-1)E_{1,m}+E_{0,m}=-J_p(L_p-1)(m-1)$. 
Hence, the zeroth order of the charge gap is $\Delta E^{(0)}=2J_p$, independent of the plaquette structure. 

\subsection{Ring plaquettes}

Here we discuss the derivation of the charge  gap for ladders of ring plaquettes [Eq.~\eqref{eq:linear_lim2}]. The zeroth order term is calculated by considering a single ring plaquette, which is treated as a chain with periodic boundary conditions. The ground state energy is $E_{1,m}=-2J_p$ if one boson is present,  and $E_{0,m}=0$ if empty, leading to  $\epsilon^{(0)}(L_p)=-2J_pL_p$ and $\epsilon^{(0)}(L_p-1)=-2J_p(L_p-1)$. The term $\epsilon^{(0)}(L_p+1)$ requires the ground state energy of a chain with periodic boundary conditions and two interacting bosons, $E_{2,m}$. The eigenstate $|\Psi\rangle$ of this system is obtained using the Bethe ansatz~\cite{wang2010preparation},
\begin{align}
|\Psi\rangle&=\sum_{x_1,x_2}\Psi(x_1,x_2)|x_1,x_2\rangle, \\
\Psi(x_1,x_2) &=A_{12}e^{i(k_1x_1+k_2x_2)}+A_{21}e^{i(k_1x_2+k_2x_1)},
\end{align}
defined in the $x_1\leq x_2$ domain, with $x_1$ and $x_2$ ($k_1$ and $k_2$) being the positions (quasimomenta) of the two bosons. By symmetry of $|\Psi\rangle$, the $x_2\leq x_1$ domain is described by $\Psi(x_2,x_1)$ and exchanged quasimomenta. Using this ansatz, the total energy and amplitudes of the system are~\cite{wang2010preparation} \begin{align}
E_{2,m}&=-2J_p\left[\cos(k_1)+\cos(k_2)\right], \\
A_{12}&=-\left(\sin{k_2}+\sin{k_1}-i\frac{U}{2J_p}\right),\\
A_{21}&=\left(\sin{k_1}+\sin{k_2}-i\frac{U}{2J_p}\right).
\end{align}
Under periodic boundary conditions, the quasimomenta obey the Bethe ansatz equations,
\begin{align}\label{eq:Bethe_1}
    \exp(ik_jm)=\frac{\sin k_l -\sin k_j -iU/(2J_p)}{\sin k_l -\sin k_j +iU/(2J_p)},
\end{align}
with $j\neq l\in[1,2]$.
Imposing the HCB limit, $U\rightarrow\infty$, leads to $k_l=\pm\pi/m$, and the two quasimomenta being either equal ($k_1=k_2$) or opposite ($k_1=-k_2$), with energy  $E_{2,m}=-4J_p\cos(\pi/m)$ in both cases.

Thus, the zeroth order of the charge gap of a ring plaquette is $\Delta E^{(0)}  = 4[1-\cos(\pi/m)]J_p$. This solution requires periodic boundary conditions which only exists for $m\geq3$. 

\subsection{First-order approximation}

We now look at the first-order perturbation, for small $J_c/J_p$, induced to $\hat{H}_{\text{plaq}}$ by $\hat{H}_{\text{conn}}$. Since $\hat{H}_{\text{conn}}$ describes inter-plaquette hopping, applying it on the zeroth order of the ground state, $|\text{GS}^{(0)}\rangle$, would result in $2(L_p-1)$ particle-hole excitations on neighboring plaquettes. The first-order energy approximation for $L_p$ bosons is $\epsilon^{(1)}(L_p) = -J_c\langle L_p|\hat{H}_{\text{plaq}}|L_p\rangle=0$. However, the first order becomes relevant for $L_p\pm1$ bosons. In the $L_p+1$ case, the ground state can be described as the superposition of many degenerate states
\begin{equation}
     |\text{GS}^{(0)}(L_p+1)\rangle=\frac{1}{\sqrt{L_p}}\sum_{j=1}^{L_p}e^{ikj}|2(j)\rangle, \nonumber
\end{equation}
with quasimomentum $k$. We define the state $|2(j)\rangle=|1\rangle\otimes\dots\otimes|2\rangle_j\otimes\dots\otimes|1\rangle$, with $L_p-1$ copies of the $|1\rangle$ state. Up to first order, $\hat{H}_\text{conn}$ only contains terms describing the transfer of a boson onto neighboring plaquettes, as higher-order-terms would be neglected. Since the inter-hopping structure is consistent between different pairs of neighbouring plaquettes, we define the $\hat{H}_\text{hop}$ matrix as connecting the $j$-th plaquette to the $(j+1)$-th one, respectively having two and one bosons. Given this, the Hilbert subspace of $\hat{H}_\text{hop}$ is of size $d(1,m)d(2,m)=m^2(m-1)/2$.
Defining $|12\rangle=|1\rangle\otimes|2\rangle$ and  $|21\rangle=|2\rangle\otimes|1\rangle$, the first-order term becomes
\begin{eqnarray}
    \epsilon^{(1)}(L_p+1)&=& -J_c\langle \text{GS}^{(0)}(L_p+1)|\hat{H}_{\text{conn}}|\text{GS}^{(0)}(L_p+1)\rangle\nonumber\\ 
    &=& -J_c\left[\langle 21| \hat{H}_\text{hop}|12\rangle e^{ik}\right.\nonumber \\ &&+\left.\langle12| \hat{H}_\text{hop}|21\rangle e^{-ik}\right] \nonumber
    \\ &=& -2S_\text{hop}J_c\cos(k),
\end{eqnarray}
where $S_\text{hop}=\sum_{i,j}(\hat{H}_\text{hop})_{i,j}$ is the sum of all the elements in the $\hat{H}_\text{hop}$ matrix.

We obtain $S_\text{hop}=C_+/[d(1,m)d(2,m)]$, where $C_+$ is the number of possible pairs of states that permit the transfer of a boson from the doubly-occupied plaquette $j$ to the singly-occupied one $j+1$. For each connection between these plaquettes, there must be a state where one site is occupied by a boson and the other is free. For that connection, there are $m-1$ states depending on the position of the second boson on plaquette $j$, multiplied by another $m-1$ accounting for the possible sites the boson on plaquette $j+1$ can occupy without blocking the hopping. This gives $C_+=(m-1)^2C$, where $C$ is the number of inter-plaquette connections, and the first-order perturbation becomes
\begin{eqnarray}
\epsilon^{(1)}(L_p+1)&=&-4J_c\frac{(m-1)}{m^2}C\cos(k). 
\end{eqnarray}

The same process is applied for the $|\text{GS}^{(0)}(L_p-1)\rangle$, but instead of $|2(j)\rangle$ we consider $|0(j)\rangle=|1\rangle\otimes\dots\otimes|0\rangle_j\otimes\dots\otimes|1\rangle$, with $S_\text{hop}=C_-/[d(1,m)d(0,m)]=C_-/m$. Here, $C_-$ is the number of pairs of states that allow for a hole to move from plaquette $j$ to $j+1$. Since we only have one particle for both plaquettes, $C_-=C$. The first-order perturbation is
\begin{eqnarray}
    \epsilon^{(0)}(L_p-1)=-\frac{2C}{m}J_c\cos(k).
\end{eqnarray}
Finally, the charge gap up to first order is
\begin{align}
    \Delta E^{(1)} &=\epsilon^{(1)}(L_p+1)+\epsilon^{(1)}(L_p-1)\nonumber \\&=-2J_c\left[\frac{3m-2}{m^2} \right]C.
\end{align}
By combining the zeroth and first-order of the charge gap, we obtain the results used in the main text in Eqs.~\eqref{eq:linear_lim} and \eqref{eq:linear_lim2}.

\bibliography{apssamp}

\end{document}